\documentclass[pre]{revtex4b3}
 
\usepackage{dcolumn}
\usepackage{amsmath}
\usepackage{amssymb}
\usepackage{epsfig}
\usepackage{multicol}          

\begin{document}

\title[Non parametric resonances of parametrically driven kinks]{Anomalies
of ac driven solitary waves with internal modes: \\
Non-parametric resonances induced by parametric forces}

\author{Niurka R.\ Quintero} 
\email{niurka@euler.us.es}

\affiliation{Grupo de F\'\i sica No Lineal, 
Departamento de F\'{\i}sica Aplicada I,  
Universidad de Sevilla,\\ Facultad de Inform\'atica,
Avenida Reina Mercedes s/n, 41012 Sevilla, Spain} 

\author{Angel S\'anchez}
\homepage{http://gisc.uc3m.es/~anxo}
 
\affiliation{Grupo Interdisciplinar de Sistemas Complicados (GISC),
Departamento de Matem\'aticas,\\ Universidad Carlos III de Madrid,
Avda.\ Universidad 30, 28911 Legan\'es, Madrid, Spain} 

\author{Franz G.\ Mertens}
\email{franz.mertens@uni-bayreuth.de}

\affiliation{Physikalisches Institut, Universit\"at Bayreuth,  
D-95440 Bayreuth, Germany}

\date{\today}

\begin{abstract} 
We study the dynamics of kinks in the $\phi^4$ model subjected to
a parametric
ac force, both with and without damping, as a paradigm of solitary
waves with internal modes. By using a collective coordinate approach,
we find that the parametric force 
has a non-parametric effect on the kink motion.
Specifically, we find that the internal mode
leads to a resonance for frequencies of the parametric driving close
to its own frequency, in which case the energy of the system 
grows as well as the width of the kink. These predictions of the 
collective coordinate theory are verified by numerical simulations of
the full partial differential equation. We finally compare
this kind of resonance with that obtained for non-parametric ac forces
and conclude that the effect of ac drivings on solitary waves with 
internal modes is exactly the opposite of their character in the 
partial differential equation.
\end{abstract}

\pacs{PACS numbers: 05.45.Yv, 02.30.Jr, 03.50.-z, 63.20.Pw}

\maketitle

\begin{multicols}{2}

\section{Introduction}
\label{intro}

During the last three decades of the 20th century, we have witnessed 
a great deal of effort devoted to the study of solitons and related 
nonlinear coherent excitations \cite{Remo,Scott,2a}.
All that work notwithstanding, the analysis of
systems generally regarded as paradigms of soliton phenomenology
still yields unknown, surprising features. One such case is the dynamics
of topological solitons or kinks in nonlinear Klein-Gordon systems 
subjected to ac forces. After several papers that led to contradictory
conclusions during the last decade, the problem for the sine-Gordon 
equation was finally solved in 1998 \cite{EPJBus}. This was the simplest
possible scenario for studying the effects of ac forces on topological
solitons, in so far as sine-Gordon kinks do not have any internal 
mode and behave as rigid objects \cite{yurirev}. However, this is a 
nongeneric situation, because many solitary waves do possess internal
degrees of freedom \cite{yurirev,Michel,David}, and even systems like
the sine-Gordon model can acquire internal modes due to the influence
of perturbations, such as discreteness \cite{Yura2,jones,kap}. Therefore, it was
necessary to study the behavior of solitons with internal modes, and 
subsequently it was found that pure, non-parametric ac forces induced 
resonances of parametric character: If the frequency of the internal 
mode is denoted by $\Omega_i$, the resonance appears when the driving
frequency is $\Omega_i/2$ \cite{prl}. This anomalous behavior was shown,
taking the $\phi^4$ equation as a typical example,
to arise from the fact that the driving couples {\em indirectly} to the 
internal mode, which in turn leads to a parametric influence on its 
evolution. 

In view of the above results, it is only natural to pose the following
question: If non-parametric drivings act parametrically on $\phi^4$ 
kinks, what is the effect of a parametric driving? To our knowledge, 
this issue was first addressed in \cite{KivSan}, although the main 
point of that work was to show how kinks can be effectively anihilated
by a fast parametric driving, and no attention was paid to the kink
dynamics. More recently, the problem was considered with both parametric
and non-parametric forces acting simultaneously in the system \cite{primak},
but the approach employed by those researchers was not appropriate and 
led to incorrect results \cite{Kiv2}. Therefore, the dynamics of 
kinks in the $\phi^4$ model
subjected to parametric drivings is largely unexplored, and this is
another reason why we concern ourselves with this problem. This paper
presents our conclusions on this question within the following scheme:
Section II is devoted to our analytical approach; Section III contains
the simulations of the full partial differential equation that verify 
our analytical predictions, and Section IV summarizes our main results.

\section{Resonances induced by parametric periodic forces}
\label{colco}

As we are interested in the behavior of solitary waves with internal 
modes, we choose as our working example the $\phi^4$ equation, which 
is well known to be representative of the generic behavior of those
solitary waves. Therefore, 
we begin by considering the perturbed problem given by
\begin{equation}
\phi_{tt} - \phi_{xx} = - \frac{dU}{d\phi} - \beta \phi_{t} + f(t,\phi),
\label{ecua1}
\end{equation}
where $U(\phi)=(1-\phi^2)^2/4$ is the $\phi^4$ potential,  
$f(t,\phi) = \epsilon \sin(\delta t + \delta_{0}) \phi$ represents
the parametric ac force,
$\epsilon$, $\delta$, and $\delta_0$ are the amplitude, frequency,
and phase of the external periodic force, respectively, and $\beta$ is the 
damping coefficient. 

In order to obtain analytical results, we resort to the well-known 
collective coordinate method \cite{yurirev,siam}, which will give us
definite predictions on the behavior of the system. However, these 
being approximate results, we will have to check them by direct 
numerical simulations of Eq.\ (\ref{ecua1}); those will be the 
subject of the following section. To apply the collective coordinate
approach to our problem, we choose to employ 
the so-called generalized 
traveling wave {\em Ansatz} \cite{franz} 
or, equivalently, to use the variation of the momentum and the
energy (the equivalence
of both techniques is proven in the second paper in \cite{prl}). These
two quantities are given by 
\begin{eqnarray}
\label{mome}
P(t) &\equiv &-\int_{-\infty}^{+\infty} dx\, \phi_{x} \phi_{t},\\
E(t) &\equiv &\int_{-\infty}^{+\infty} dx\left\{
\frac 12 \phi_t^2+\frac12\phi_x^2+U(\phi)\right\}.
\label{hami}
\end{eqnarray}
In order to evaluate those quantities, 
we consider the Rice {\em Ansatz} \cite{Rice} 
and assume that 
\begin{equation}
\phi(x,t) = \phi_0\left[\frac{x-X(t)}{l(t)}\right], 
\end{equation}
where $\phi_0(x)$ is the static kink solution of the
unperturbed $\phi^4$ equation, i.e., of Eq.\ (\ref{ecua1}) 
with $\beta=\epsilon=0$ zero, centered at $X(t)$ and 
with width $l(t)$. With this choice, and following the 
standard procedure \cite{yurirev,siam},
we find that the evolution of the momentum, 
$P(t) \equiv M_{0} l_{0} \dot{X}/l(t)$, and the width of 
the kink are given by 
\begin{eqnarray}
\frac{dP}{dt} & = & F^{stat}(X) -  \beta P + F_{ex}, 
\label{evmo} \\
\alpha M_{0} l_{0} \frac{\ddot l}{l} + \frac{P^{2}}{M_{0} l_{0}} 
& = & K^{int}(l,\dot{l},\dot{X}) -  \beta \alpha M_{0} l_{0} \frac{\dot l}{l} + K, \label{evl} 
\end{eqnarray}
with  
\begin{eqnarray} 
F_{ex} & = & \int_{-\infty}^{+\infty} dx \,\ f(t,\phi) 
\frac{\partial \phi}{\partial X}, \,\ \,\ F^{stat}= -\frac{\partial E}
{\partial X},  \\
K & = & \int_{-\infty}^{+\infty} dx \,\ f(t,\phi) \,\ 
\frac{\partial \phi}{\partial l}, \,\ \,\ 
K^{int} = -\frac{\partial E}{\partial l},   \\ 
E & = & \frac{1}{2} \frac{l_{0}}{l} M_{0}  \dot{X}^{2} + 
\frac{1}{2} \frac{l_{0}}{l} \alpha M_{0}  \dot{l}^{2} + 
\frac{1}{2} M_{0} \left(\frac{l_{0}}{l} + \frac{l}{l_{0}}\right), 
\label{ener}
\end{eqnarray}
where 
$q=2$, $M_{0}=4/(3 l_{0})$, and $l_{0}=\sqrt{2}$ represent the
topological charge, the mass, and the width of the unperturbed kink, 
respectively, and 
$\alpha=(\pi^2-6)/12$ is a constant.

We now have to evaluate the above quantities. In this respect,
it is interesting to compare 
the equations we have obtained with those found for the 
case of an ac non-parametric force, i.e. $f(t,\phi)=f(t)$. In that situation
we obtain that  
$F_{ex} = - q f(t)$ and $K=0$, and hence
the external force acts directly on the  
translational mode, whereas the internal mode is excited indirectly
due to the coupling between these two modes. 
Conversely, as we have seen,
in the parametrically driven problem, $F_{ex}$ vanishes, while 
$K = - \epsilon \sin(\delta t + \delta_{0})$. Therefore, Eqs.\ 
(\ref{evmo})-(\ref{evl}) become 
\begin{eqnarray}
\frac{dP }{dt} &=& -  \beta P(t),    
\label{ecua23}\\
\alpha \left[\dot{l}^{2} - 2 l \ddot{l} - 2 \beta l \dot{l}\right] & = &  
\frac{l^{2}}{l_{0}^{2}} \left [1 +
\frac{P^{2}}{M_{0}^{2}} \right ] \nonumber\\ &-& 1 + 
2 \frac{l(t)^2}{M_{0} l_{0}}\, \epsilon \sin(\delta t + \delta_{0}).
\label{ecua24}
\end{eqnarray}
We thus see that, when introduced in a parametric manner,
the ac force acts directly on the internal mode and
not via the coupling between the translational and internal 
modes. This result shows the very different role of both drivings
at the collective coordinate level. 

In order to deal with these equations, we choose
${\dot X}(0)=u(0)$, $l(0)\equiv l_{s} = l_{0} \sqrt{1-u^2(0)}$ 
and $\dot{l}(0)$ as the initial conditions for Eqs.\ (\ref{ecua23}) and 
(\ref{ecua24}). 
The equation for the momentum is trivial and 
can be solved exactly, yielding 
\begin{eqnarray}
P(t) & = & P(0) {\rm e}^{-\beta t}.   
 \label{ecua4}
\end{eqnarray} 
We note that for $\beta \ne 0$, the momentum 
of the kink goes to zero and after some transient time, $t >> 1/\beta$, 
it effectively vanishes.  In the same way,
if we start from zero initial velocity, $P(t)=0$, i.e., the 
center of the 
kink will not move.

Let us now turn to Eq.\ (\ref{ecua24}). As a first step, we
introduce a change of variables, proposed 
in \cite{yuri}, given by $l(t)=g^{2}(t)$.
This change transforms Eq.\ (\ref{ecua24}) into 
an Ermakov-type equation (or Pinney-type) \cite{Pinney}, which reads
\begin{eqnarray}
\label{ecua25}
\ddot{g} &+& \beta \dot{g}  +
\Big[ \left( \frac{\Omega}{2} \right)^2 \\ &+&
\left( \frac{\Omega}{2 M_{0}} \right)^2 P^2 + 
\epsilon \frac{\sin(\delta t + \delta_{0})}{2 \alpha M_{0} l_{0}} 
\Big] g =  
\frac{1}{4 \alpha g^{3}}, \nonumber \\
 g(0) &=& \sqrt{l_{s}} \ne 0,   \,\ \,\ \,\   
\dot{g}(0) = \frac{\dot{l}(0)}{2 \sqrt{l_{s}}},
\end{eqnarray}
where $\Omega = 1/\sqrt{\alpha}  l_{0}=1.2452$ is equal to   
the Rice frequency \cite{Rice} $\Omega_{R} = 1/\sqrt{\alpha}  l_{s}$ 
in the case when the kink initially is at rest.  
It can be shown (see \cite{Rice} and \cite{epjb3}) that when there is
not any perturbation in the system (\ref{ecua23})-(\ref{ecua24}),  
$l(t)$ oscillates with a frequency
$\Omega_R$ if we start from any $l(0)\neq l_s$ or $\dot{l}(0)\neq0$. Therefore,
$\Omega_R$ is the characteristic frequency of Eq.\ (\ref{ecua24}) and,
since it agrees within $1.7\%$ with 
$\Omega_i=\sqrt{3/2}=1.2247$,
we expect that if we find a resonance related with $\Omega_R$
in Eq.\ (\ref{ecua25}) [or Eq.\ (\ref{ecua24})] we should find a similar 
phenomenon in the full system, Eq.\ (\ref{ecua1}),
associated to the frequency of the internal mode 
$\Omega_i$.

We now
proceed to solve the Eq.\ (\ref{ecua25}) analytically for the 
undamped case, i.e., when $\beta=0$.
As we have mentioned, in this case $P(t)=P(0)$ and 
Eq.\ (\ref{ecua25}) becomes the following Pinney-type 
equation \cite{Pinney}  
\begin{equation}
\ddot{g} + 
\left [ \left(\frac{\Omega_{R}}{2}\right)^2 + 
\epsilon \frac{\sin(\delta t + \delta_{0})}{2 \alpha M_{0} l_{0}} 
\right ] g =  
\frac{1}{4 \alpha g^{3}}, 
\label{ecua26}
\end{equation}
whose solution is 
\begin{equation}
\begin{array}{c}
{\displaystyle{
g(t)  =  \sqrt{v_{1}^{2} + \frac{1}{4 \alpha W^{2}} v_{2}^{2}}}}, 
\end{array}
\label{ecua27}
\end{equation}
where $v_{1}(t)$ and $v_{2}(t)$ are two independent solutions of the linear
part of Eq.\ (\ref{ecua26}) and $W=\dot{v}_{1} v_{2} - \dot{v}_{2} v_{1}$ 
is the Wronskian. $W(t)$ is actually a constant, 
and can be calculated from 
the initial conditions for
$v_{i}$ ($i=1,2$), $v_{1}(0)=\sqrt{l_{s}}$,  $\dot{v}_{1}(0) 
= \dot{l}(0)/(2 \sqrt{l_{s}})$, 
$v_{2}(0) = 0$ and  $\dot{v}_{2}(0)$ a nonzero constant. 

If one denotes $\tau =(\delta t + \delta_{0} + \pi/2)/2$, 
after some algebraic 
manipulations we arrive at the following Mathieu 
equation for the $v_{i}$ functions:
\begin{eqnarray}
v_{i}'' &+& [a - 2 \theta \cos(2 \tau) ] v_{i} = 0,  \nonumber \\
a=\left(\frac{\Omega_{R}}{\delta}\right)^2, & &
\theta = \frac{\epsilon}{\alpha M_{0} l_{0} \delta^2} \equiv 
\frac{\epsilon l_{0}}{M_{0} \gamma_{0}^{2}} 
\left(\frac{\Omega_{R}}{\delta}\right)^{2}, 
\label{ecua28}
\end{eqnarray}
where prime denotes the derivative with respect to $\tau$,
and $\gamma_0=1/\sqrt{1-v(0)^2}$. 
Notice that the initial  conditions, when $\tau \equiv \tau_{0}= 
(\delta_{0} + \pi/2)/2$ for $v_{i}(\tau)$, become 
$v_{1}(\tau_0)=\sqrt{l_{s}}$,  
$v_{1}'(\tau_0)= \dot{l}(0)/(\delta \sqrt{l_{s}})$,  
$v_{2}(\tau_0) = 0$ and $v_{2}'(\tau_0)= 2 \dot{v}_{2}(0)/\delta$. 
The solution of Eq.\ (\ref{ecua28}) (see \cite{Mathieu}) for $v_{1}(\tau)$ and
$v_{2}(\tau)$ can be expressed as  a linear  superposition of the two Mathieu
functions ${\rm{ce}}_{\nu}$ and ${\rm{se}}_{\nu}$ 
with a non-integer index $\nu$, i.e., 
\begin{equation}
\begin{array}{l}
{\displaystyle{
v_{i}(\tau) = A_{i} \, {\rm {ce}}_{\nu}\left(\tau,\theta\right) + 
B_{i} \, {\rm{se}}_{\nu}\left(\tau,\theta\right), \,\ \,\ i=1,2,}}  
\end{array}
\label{ecua29}
\end{equation}
where
\begin{equation}
\begin{array}{l}
{\displaystyle{
 A_{i}  \equiv  \frac{\Delta_{A_{i}}}{\Delta}, \,\ \,\ 
 B_{i} \equiv \frac{\Delta_{B_{i}}}{\Delta}}},  
\label{ecua30} 
\end{array}
\end{equation}
and
\[
\begin{array}{l}
{\displaystyle{
\Delta = {\rm {ce}}_{\nu}\left(\tau_{0},\theta\right) \, 
 {\rm {se}}'_{\nu}\left(\tau_{0},\theta\right) \, - \,
{\rm {ce}}'_{\nu}\left(\tau_{0},\theta\right) \,
 {\rm {se}}_{\nu}\left(\tau_{0},\theta\right)}}, \nonumber \\[3mm] 
{\displaystyle{
\Delta_{A_{i}} = v_{i}(\tau_{0}) \, 
{\rm {se}}'_{\nu}\left(\tau_{0},\theta\right) \, - \,  
v'_{i}(\tau_{0}) \,
 {\rm {se}}_{\nu}\left(\tau_{0},\theta\right)}}, \nonumber \\[3mm]
{\displaystyle{
\Delta_{B_{i}} = v'_{i}(\tau_{0}) \,  
{\rm {ce}}_{\nu}\left(\tau_{0},\theta\right) \, - \,  
v_{i}(\tau_{0}) \, {\rm {ce}}'_{\nu}\left(\tau_{0},\theta\right)}} 
, \nonumber
\end{array}
\]
with the constraint (characteristic curve for Mathieu functions)
\begin{equation}
\begin{array}{l}
{\displaystyle{
a = \nu^{2} + \frac{1}{2 (\nu^{2}-1)} \theta^{2} + O(\theta^{4})}}. 
\end{array}
\label{ecua31}
\end{equation}

{}From Eqs.\ (\ref{ecua27}), (\ref{ecua29}) and (\ref{ecua30}), and taking
into  account that $\tau = (\delta t + \delta_{0} + \pi/2)/2$ 
we obtain that the kink width $l(t)$
is given by 
\begin{equation}
\begin{array}{l}
{\displaystyle{
l(t) = g^{2} = v_{1}^{2}(t) + \frac{1}{4 \alpha W^{2}} v_{2}^{2}(t)}}, 
\end{array}
\label{ecua32}
\end{equation}
where
\begin{eqnarray}
v_{i}(t) &=& A_{i} \,  
{\rm {ce}}_{\nu}\left(\delta t/2 + \delta_{0}/2 + \pi/4,\theta\right) 
\nonumber \\
\label{ecua33}
&+&B_{i} \, {\rm{se}}_{\nu}\left(\delta t/2 + \delta_{0}/2 + \pi/4,
\theta\right), \,\ \,\ i=1,2;\\
\label{ecua35}
W &=& -\sqrt{l_{s}} \dot{v}_{2}(0),    
\end{eqnarray}
and the characteristic curve, Eq.\ (\ref{ecua31}),
for our initial parameters can be written  
up to order $\epsilon^{2}$ as  
\begin{equation}
\delta = \frac{\Omega_{R}}{\nu} \left(1 - 
\frac{\nu^{2} l_{0}^{2}}{4 (\nu^2-1) M_{0}^{2} \gamma_{0}^4} 
\epsilon^{2}  \right) + O(\epsilon^{4}).  
\label{ecua36}
\end{equation}

\begin{figure}
\epsfig{figure=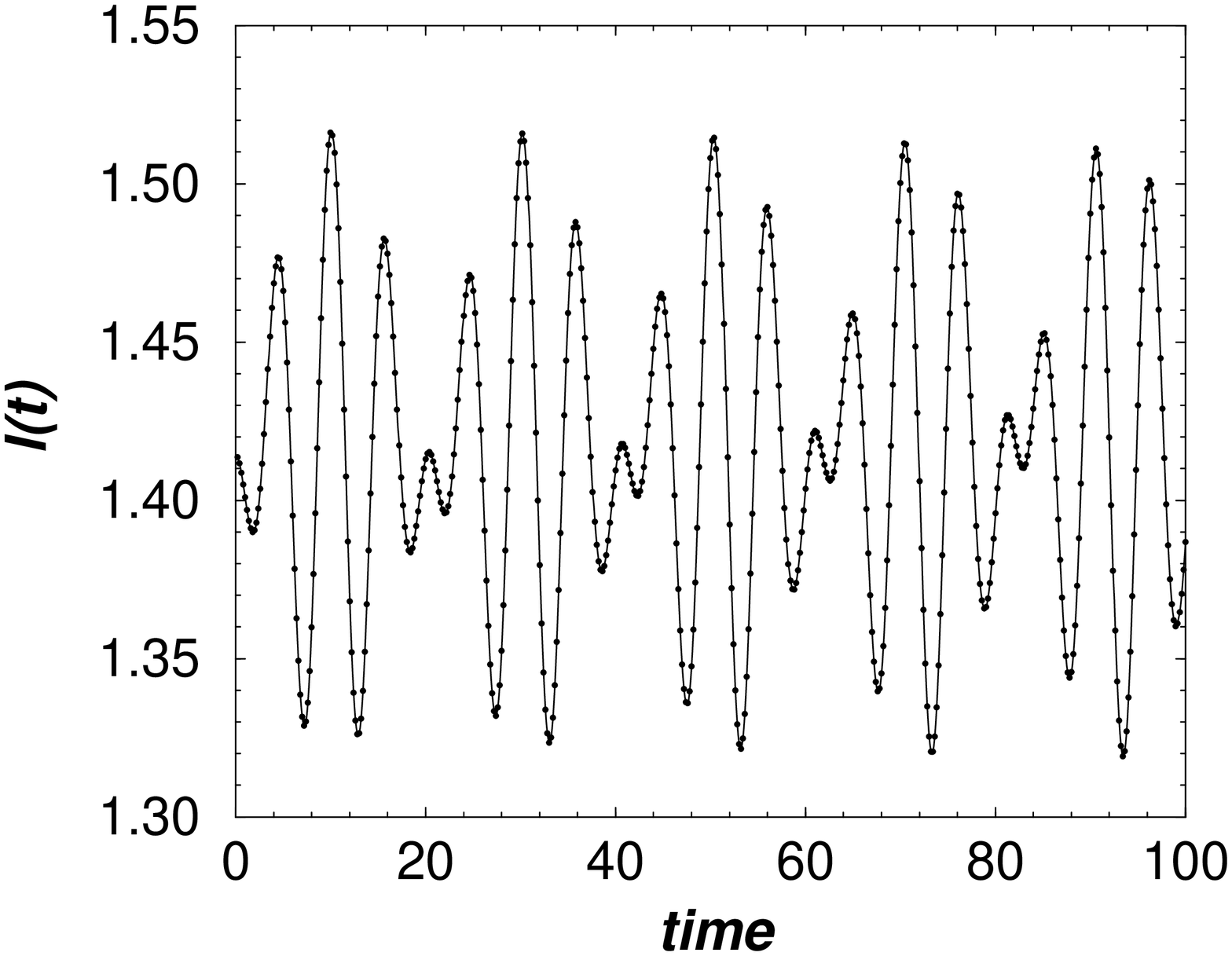, width=2.3in, angle=-90} \\
\\
\epsfig{file=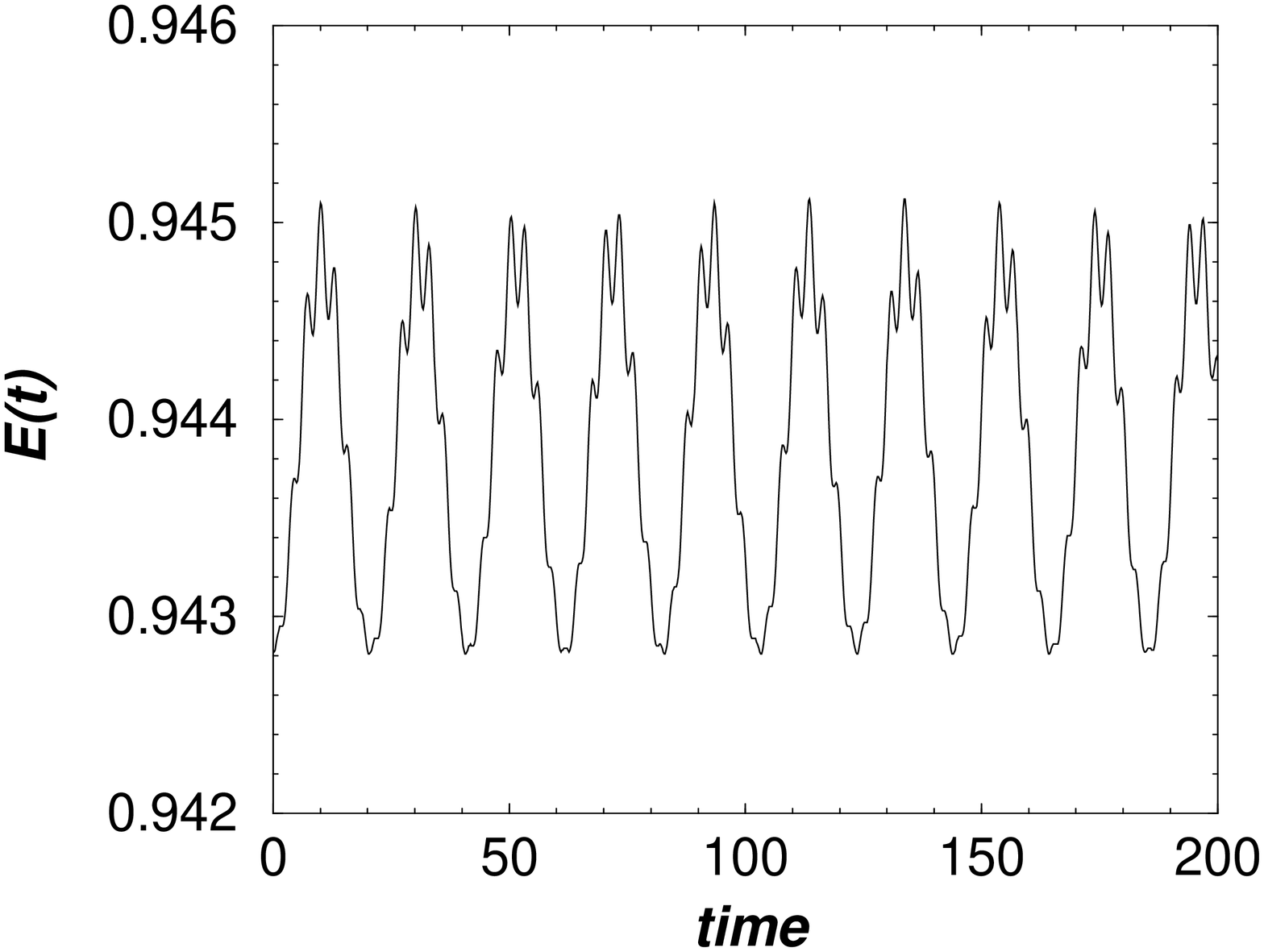,width=2.23in,angle=-90} 
\caption{Collective coordinates: 
Evolution of $l(t)$ and $E(t)$ when $\delta=0.94$ for the undamped case,
$\beta=0$. 
Points: Analytical solution [see Eqs.\ 
(\ref{ecua29})-(\ref{ecua35})]. 
Solid lines: Numerical integration of Eq.\ (\ref{ecua24}) 
for $\beta=0$, $\epsilon=0.01$, $\delta_{0}=\pi/2$, $u(0)=0$, $X(0)=0$. 
The DFT (not shown) of $l(t)$ yields two frequencies: 
$\omega_{1}=0.9434 \approx \delta$ 
and $\omega_{2}=1.25789 \approx \Omega_{R}$, whereas for
$E(t)$ it gives
three frequencies: 
$\omega_{1}=0.3144 \approx |\delta - \Omega_{R}|=0.3052$, 
$\omega_{2}=1.8868 \approx 2 \delta=1.88$ and 
$\omega_{3}=2.20131 \approx \delta + \Omega_{R}=2.1852$ .
The maximum of the 
DFT appears at $\omega_{1}$.}
\label{f1}
\end{figure}

Notice that when $\nu=m+p/s$ is rational, 
with $m$ an integer number and 
$p/s$ a rational fraction ($0 < p/s < 1$),  
 $v_{1}(t)$ and $v_{2}(t)$ are $2 \pi s$ periodic functions, if $p$ is odd,  
and $\pi s$-periodic functions, if $p$ is even,
whereas for irrational $\nu$ both functions will 
be non-periodic although bounded \cite{Mathieu}. 
For instance, if we take in Eq.\ (\ref{ecua24}) $\epsilon =0.01$, 
$\delta_0=\pi/2$, $u(0)=0$,
$l(0)=l_s$ and $\delta=0.94$, we can see in Fig.\ \ref{f1} 
that $l(t)$ is a bounded function. Moreover, taking the Mathieu 
functions that appear in Eq. (\ref{ecua33}) up to terms of order 
$\theta$ and substituting the approximate expression
of Eq.\ (\ref{ecua33}) in Eq.\ (\ref{ecua32})
one can show that the resulting 
expression for $l(t)$ involves basically the frequencies 
$\Omega_R$ and $\delta$ (see Fig. \ref{f1}).
We have also plotted in Fig.\ \ref{f1} the evolution of
the energy, Eq.\ (\ref{ener}),
taking $\delta$ not close to $\Omega_R$. For this 
choice of parameters, $E(t)$
is also a bounded function and its spectrum for the above parameters
involves chiefly three frequencies: $|\delta-\Omega_{R}|$, $2 \delta$ 
and $\delta + \Omega_{R}$.   

\begin{figure}
\epsfig{file=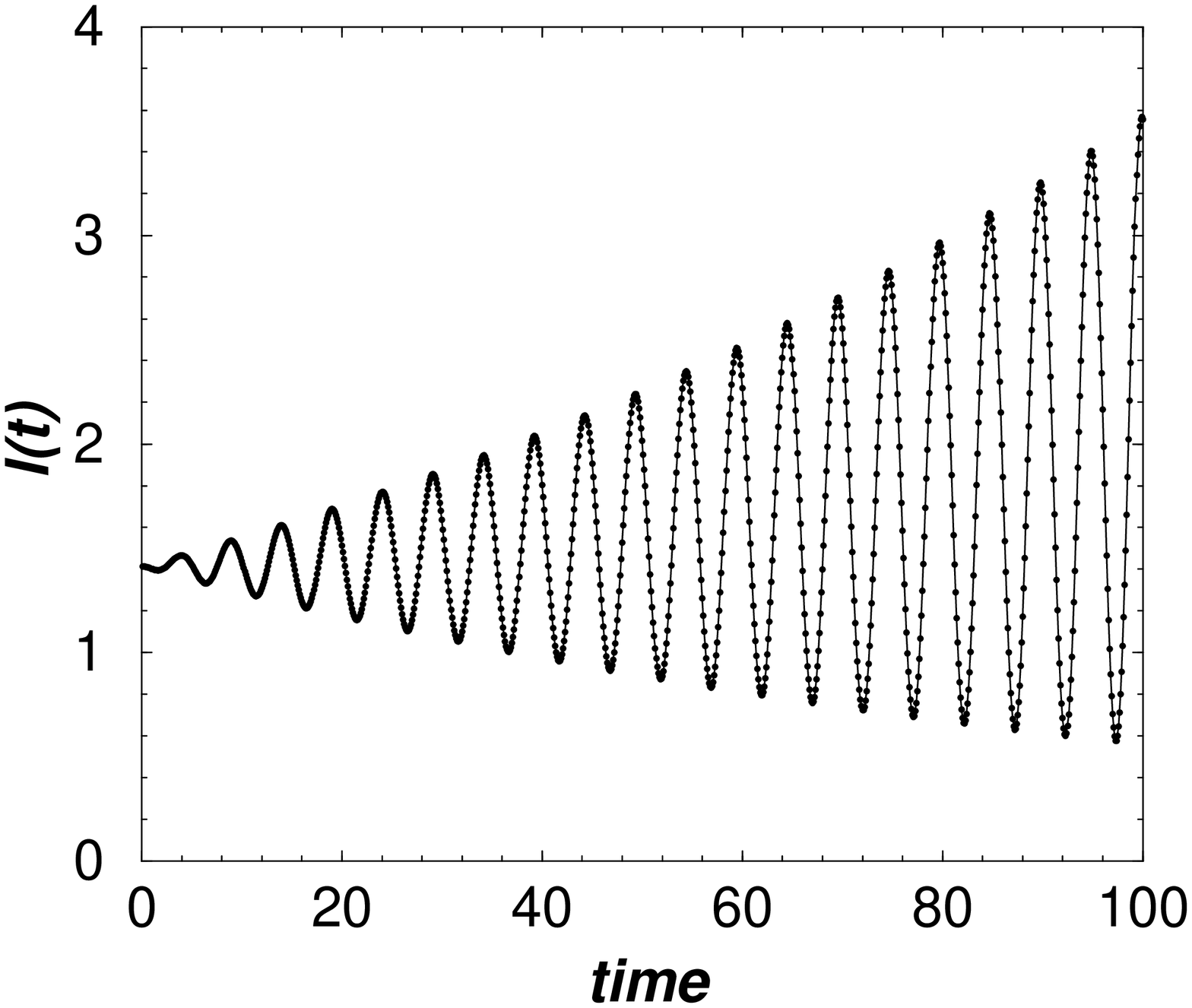,width=2.3in,angle=-90} \\
\\
\epsfig{file=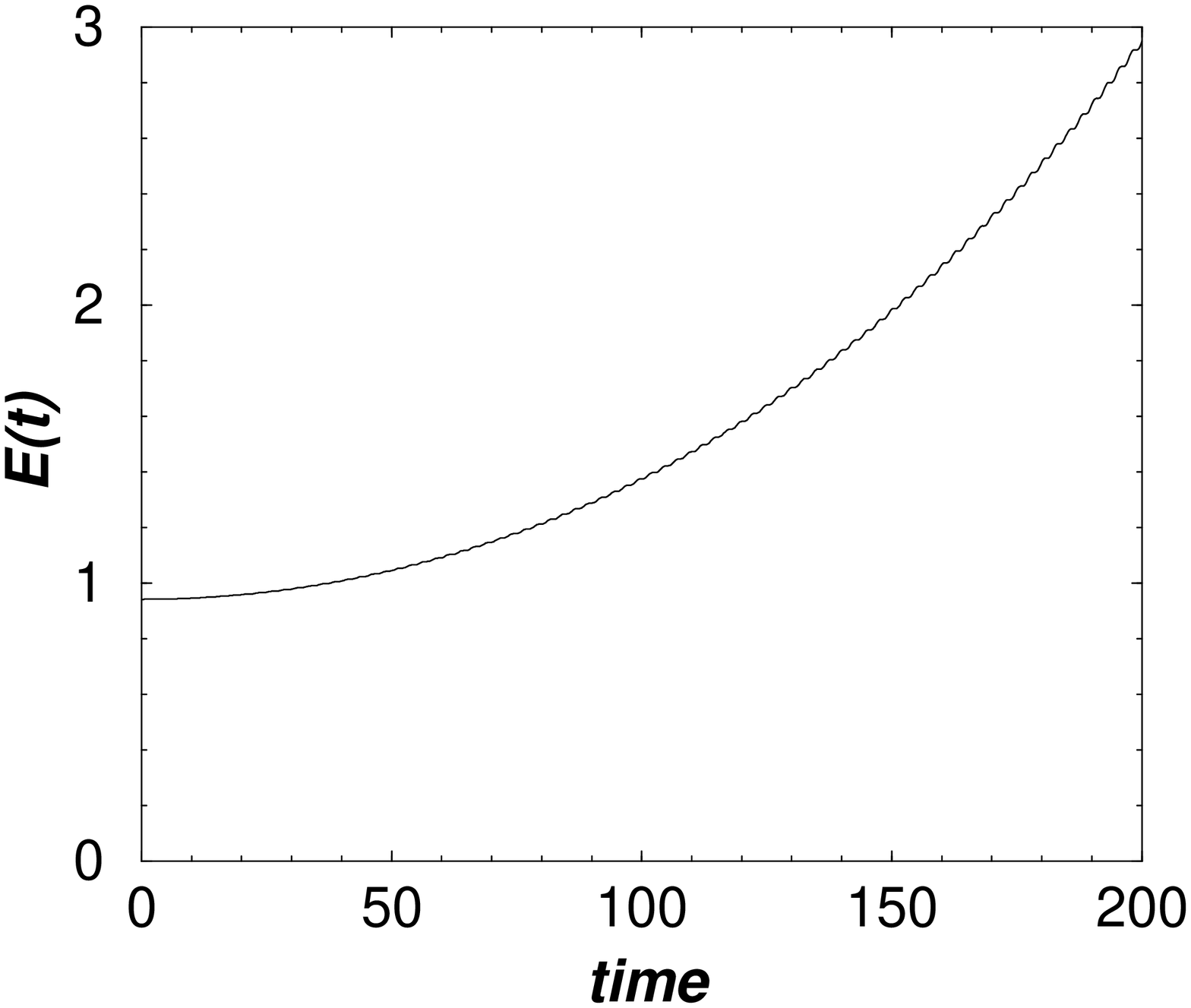,width=2.3in,angle=-90} 
\caption{Collective coordinates: 
Evolution of $l(t)$ and $E(t)$ when $\delta=1.24\approx \Omega_R$
for the undamped case, $\beta=0$.
Points: Analytical solution [see Eqs.\ 
(\ref{ecua30})-(\ref{ecua35})]. 
Solid lines: Numerical integration of Eq.\ (\ref{ecua24}). 
Parameters are as in the previous figure. 
}
\label{f2}
\end{figure}

As we have shown, Eq.\ (\ref{ecua36}) represents 
the characteristic curve corresponding to stable (either periodic or 
bounded) solutions of the Mathieu equation (\ref{ecua28}).
However, if we try to find a periodic solution when 
$\delta \approx \Omega_{R}$, we obtain that the integer Mathieu functions 
${\rm se}_{1}$ and ${\rm ce}_{1}$ are two independent solutions of two 
different Mathieu equations since they are related to different 
characteristic curves: 
\begin{eqnarray}
a &=& 1 - \theta - \frac{\theta^{2}}{8} + \frac{\theta^{3}}{64} - 
\frac{\theta^{4}}{1536} + \ldots,\\
a &=& 1 + \theta - \frac{\theta^{2}}{8} - \frac{\theta^{3}}{64} - 
\frac{\theta^{4}}{1536} + \ldots
\end{eqnarray}
respectively. These characteristic curves separate the unstable and 
stable regions of Eq.\ (\ref{ecua28}). 
In the unstable regions the solution of this equation, as well as 
$l(t)$ and $E(t)$, grow with $t$ (see Fig.\ \ref{f2}). 
This is hence the hint on the existence of a resonance phenomenon,
and,
as we already pointed out, we expect that it will manifest itself
in the full partial differential equation with an increment of
the width and the energy of the kink, when $\delta$ is close
to $\Omega_i$. This is the aim of the next section, in which we 
we will check by computing the solution of (\ref{ecua1}) whether
there 
is some resonance at $\delta\approx\Omega_i$. In addition, 
we will analyze 
numerically the effect of damping in our resonance picture, as we 
have not been able to obtain any analytical approximate result
in that case.  

\section{Simulation results}

\label{numso}

In this section we carry out numerical simulations 
of Eq.\ (\ref{ecua1}) by using the Strauss-V\'azquez scheme
\cite{stvaz},
taking the length of the system $L=100$, $\Delta t=0.01$,
$\Delta x=0.1$, the final time $t_f=200$ and free boundary 
conditions. We fix also the amplitude $\epsilon =0.01$ and the phase
$\delta_0=\pi/2$ of the ac force. 

\begin{figure}
\epsfig{figure=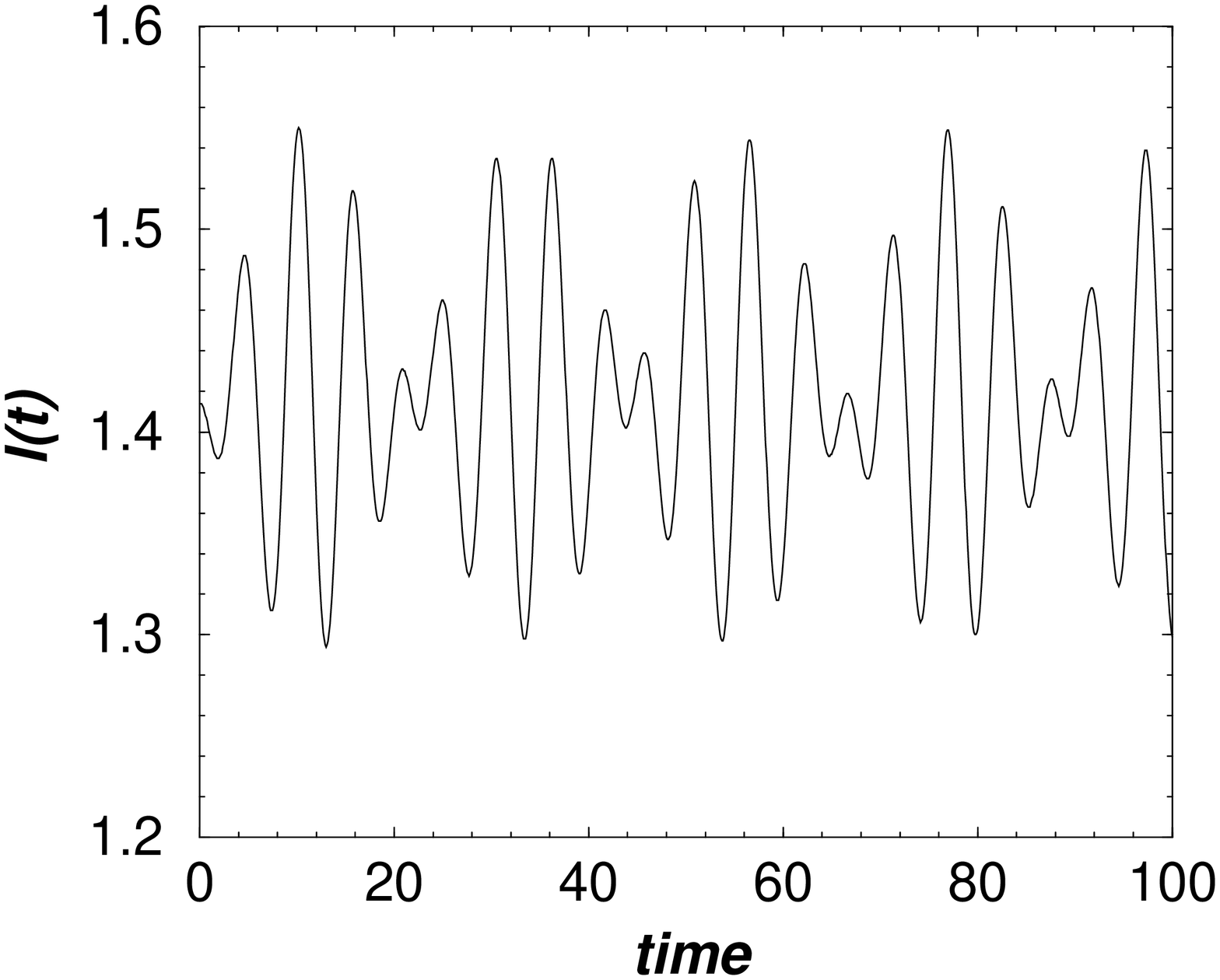, width=2.3in, angle=-90} \\
\\
\epsfig{file=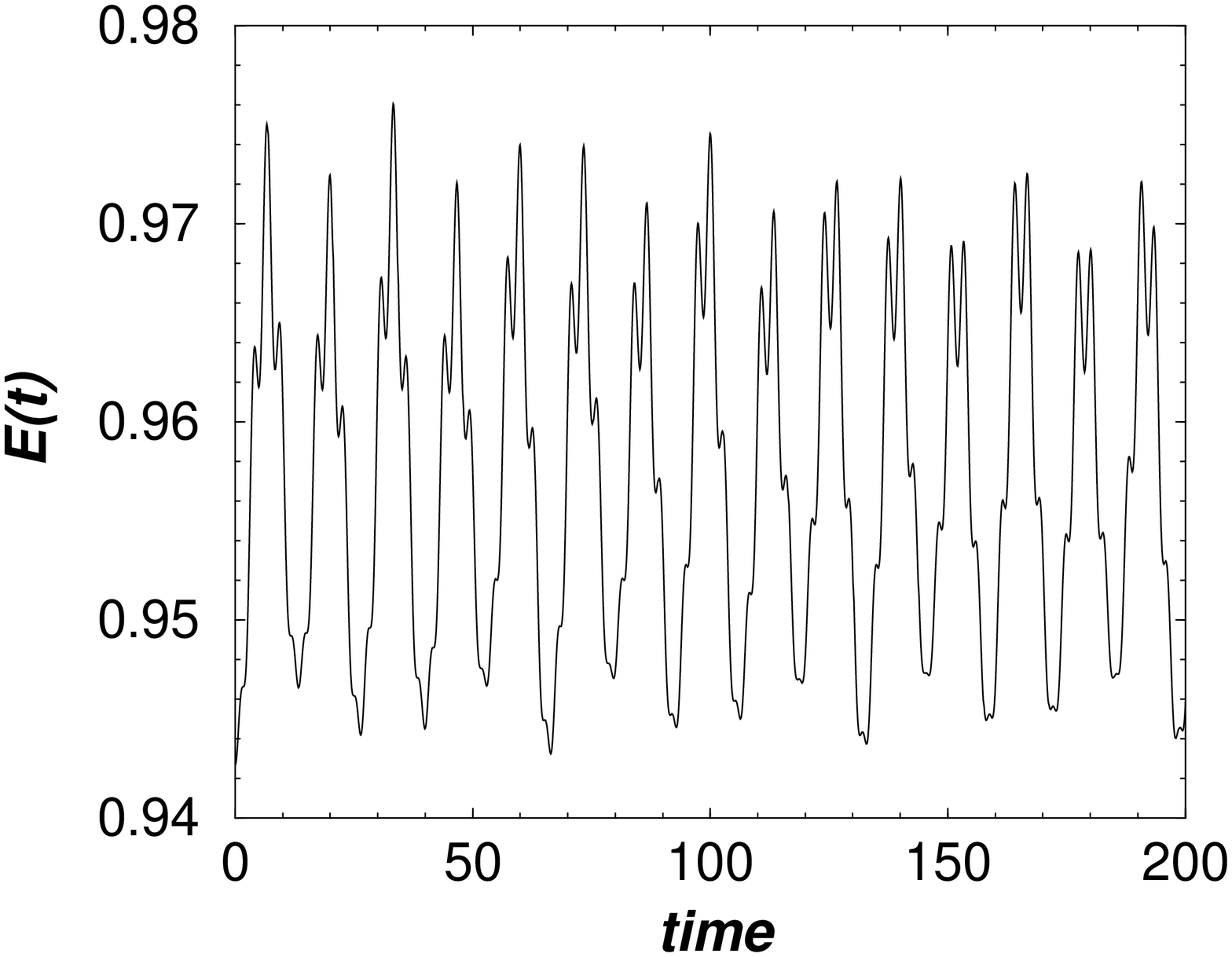,width=2.22in,angle=-90} 
\caption{
Evolution of $l(t)$ and $E(t)$ as obtained from numerical 
simulations of Eq.\ (\ref{ecua1}), when $\delta=0.94$ with  
$\epsilon=0.01$, $\delta_{0}=\pi/2$, $u(0)=0$, $X(0)=0$. 
The DFT (not shown) of $l(t)$ yields two frequencies: 
$\omega_{1}=0.9429 \approx \delta$ 
and $\omega_{2}=1.2258 \approx \Omega_{i}$, whereas for 
$E(t)$ 
gives four frequencies: 
$\omega_{1}=0.2887 \approx |\delta - \Omega_{i}|=0.2847$, 
$\omega_{2}=1.8859 \approx 2 \delta = 1.88$,   
$\omega_{3}=2.3574 \approx \delta+\Omega_{i}=2.1647$ and 
$\omega_{4}=0.4715 \approx \omega_{0}-\delta=0.4749$.   
}
\label{f3}
\end{figure}

In order to show the evolution of $l(t)$ and $E(t)$ in the 
PDE (\ref{ecua1}) we distinguish two cases: In the first case we 
choose $\delta$ far away from $\Omega_i$, whereas in the second one
$\delta\approx\Omega_i$. In Fig.\ \ref{f3} we plot the width and the 
energy of the kink as functions of $t$ for $\delta=0.94$, i.e., 
the parameters of the ac force are the same as in Fig.\ \ref{f1}. If we
compare Figs.\ \ref{f1} and \ref{f3} we see that $l(t)$ and 
$E(t)$ have the same behavior both on the collective coordinate level
and in the full system, Eq.\ (\ref{ecua1}).
Naturally, while in 
the spectrum of $l(t)$ obtained by the collective coordinate approach
there appear the frequencies $\delta$ and $\Omega_R$, in the DFT of $l(t)$ 
for the PDE we obtain $\delta$ and $\Omega_i$ as expected. However, in the 
Fourier transform of $E(t)$ for the full system we find not only frequencies 
which involve $\delta$ and $\Omega_{i}$ (see Fig.\ \ref{f3}), but also 
an additional frequency, $\omega_{4}=0.4715 \approx \omega_{0} - \delta$, 
related with the frequency of the lowest phonon 
($\omega_{0} = \sqrt{2}$) of the $\phi^4$ system. This means that the 
internal mode and the phonons can both appear when the 
$\phi^4$ system is driven parametrically with an ac force, but since 
$\Omega_{i}$ and $\omega_{0}$ are separated we expect to be able to 
excite the internal mode to a larger extent 
than phonon modes if the frequency of the 
ac force is closer to $\Omega_{i}$ than to $\omega_0$.

\begin{figure}
\epsfig{file=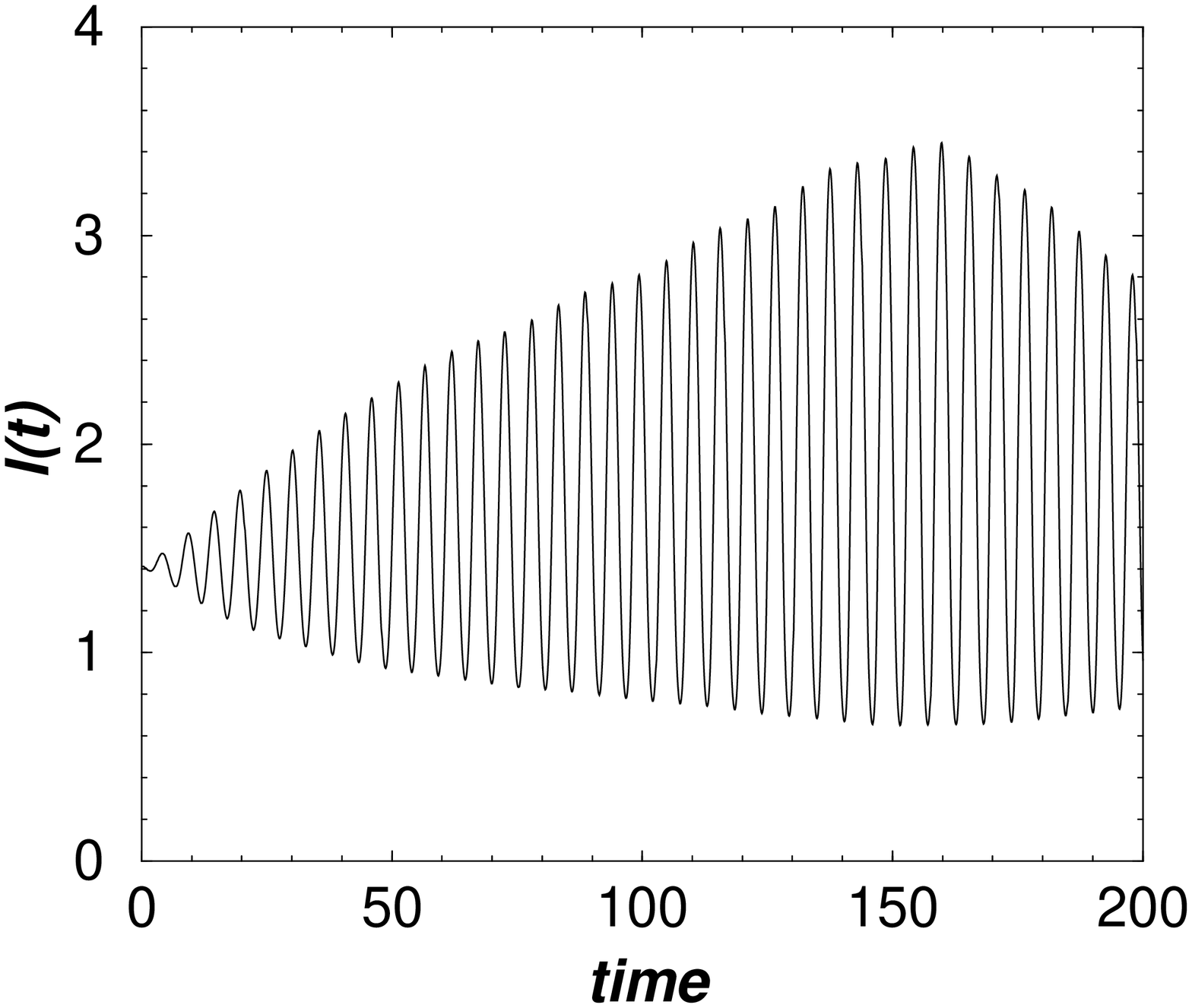,width=2.3in,angle=-90} \\
\\
\epsfig{file=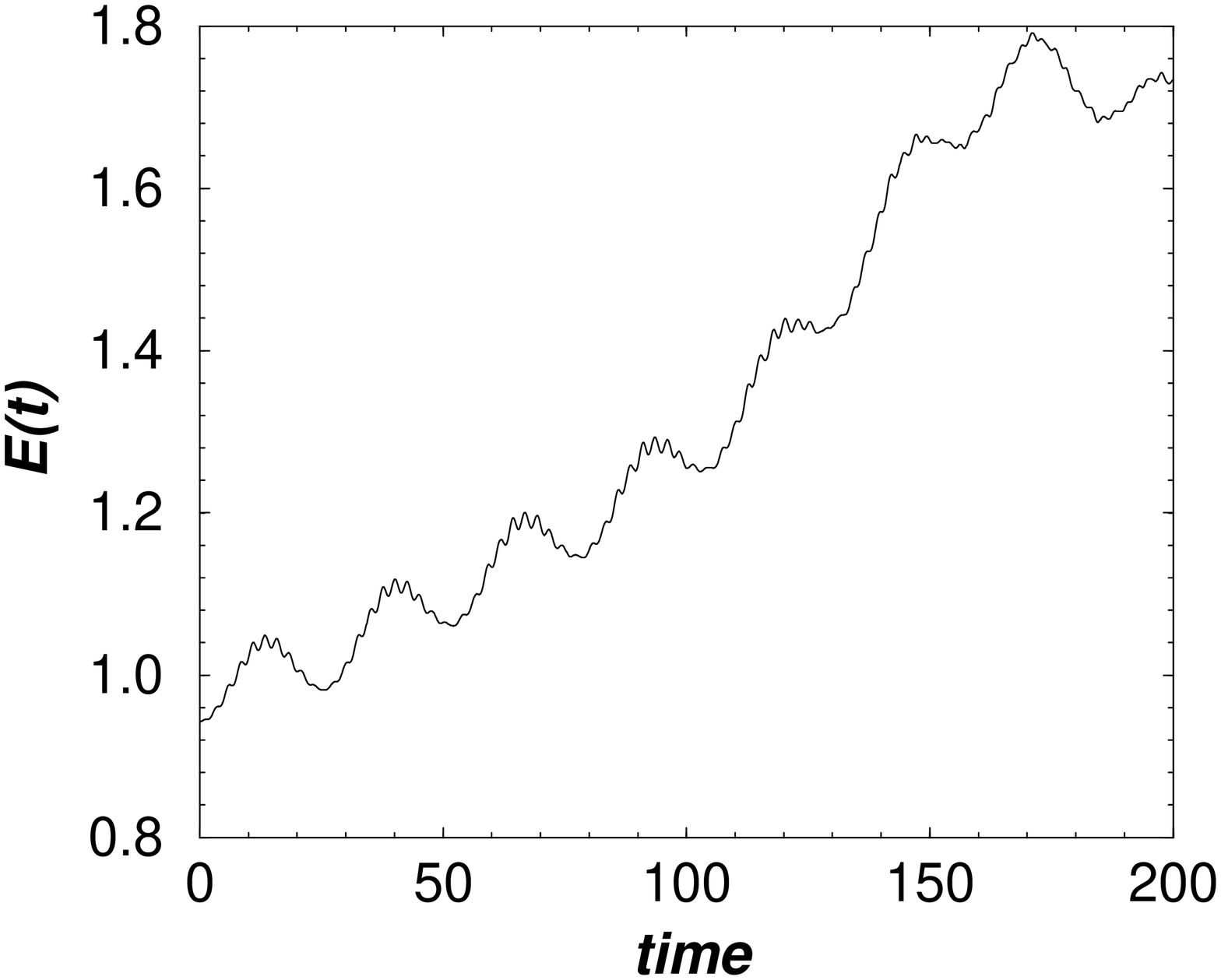,width=2.22in,angle=-90} 
\caption{
Evolution of $l(t)$ and $E(t)$ as obtained from numerical 
simulations of Eq.\ (\ref{ecua1}), when $\delta=1.176$.
Other parameters are the same as in the previous figure. 
}
\label{f4}
\end{figure}

Moving now to the second case, i.e., drivings with frequency 
$\delta$ close to $\Omega_{i}$, 
Fig.\ \ref{f4} shows that in this situation
the width of the kink oscillates with a large amplitude 
while the energy 
grows, although not monotonously. 
It is important to stress at this point
that, whereas the collective coordinate analysis predicts that
$l(t)$ and $E(t)$ 
should increase indefinitely at resonance,  
this is not the case in the original system. 
The reason is that, in our perturbative method, we have not
taken into 
account the phonons, which are present in the system, 
and therefore part of the 
energy input from the ac force goes to the phonons. 
In our simulations we have also 
seen that the kink does not move, even if we 
are at resonance, in agreement with our collective
coordinate theory.

\begin{figure}
\epsfig{figure=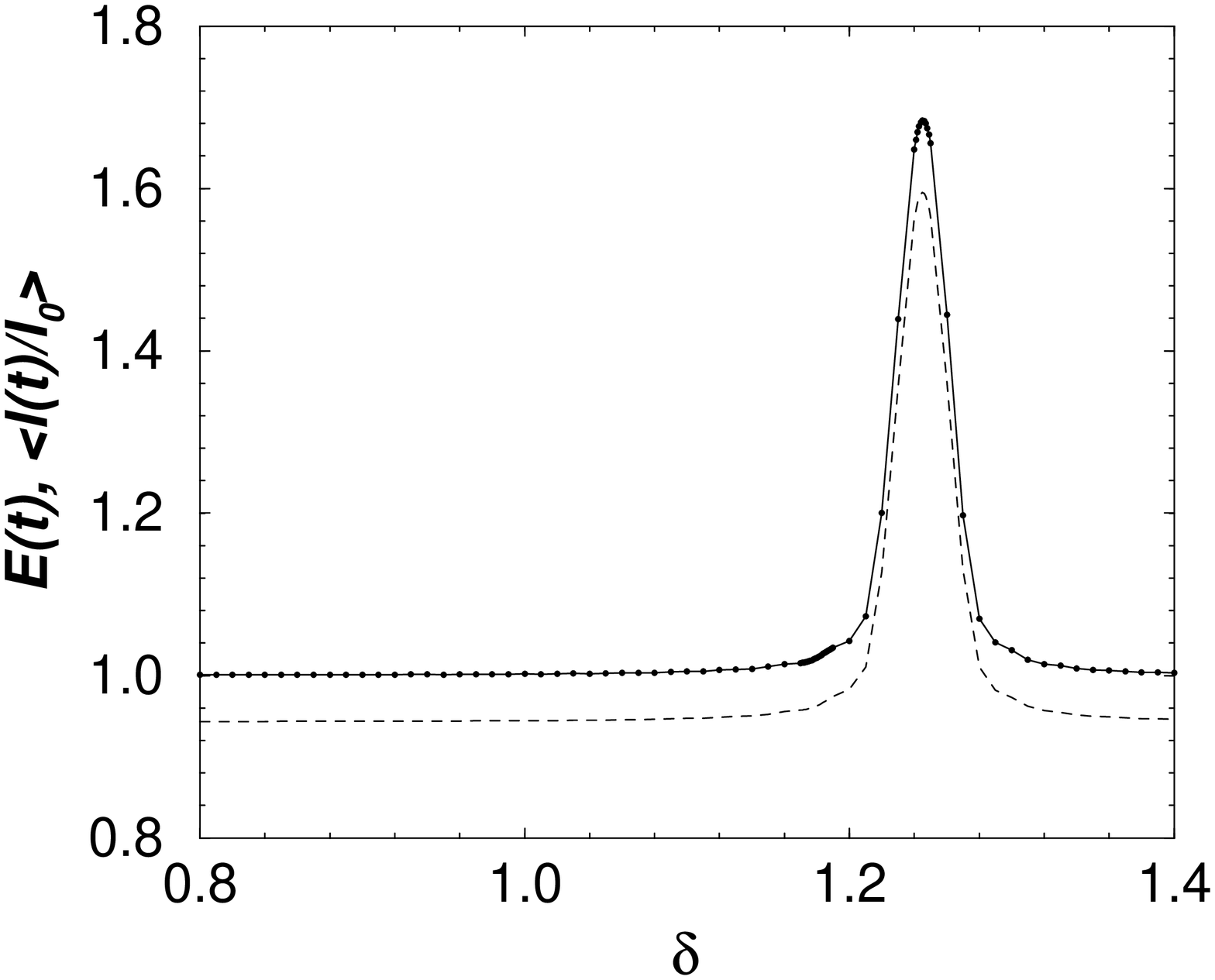, width=2.3in, angle=-90} \\
\epsfig{figure=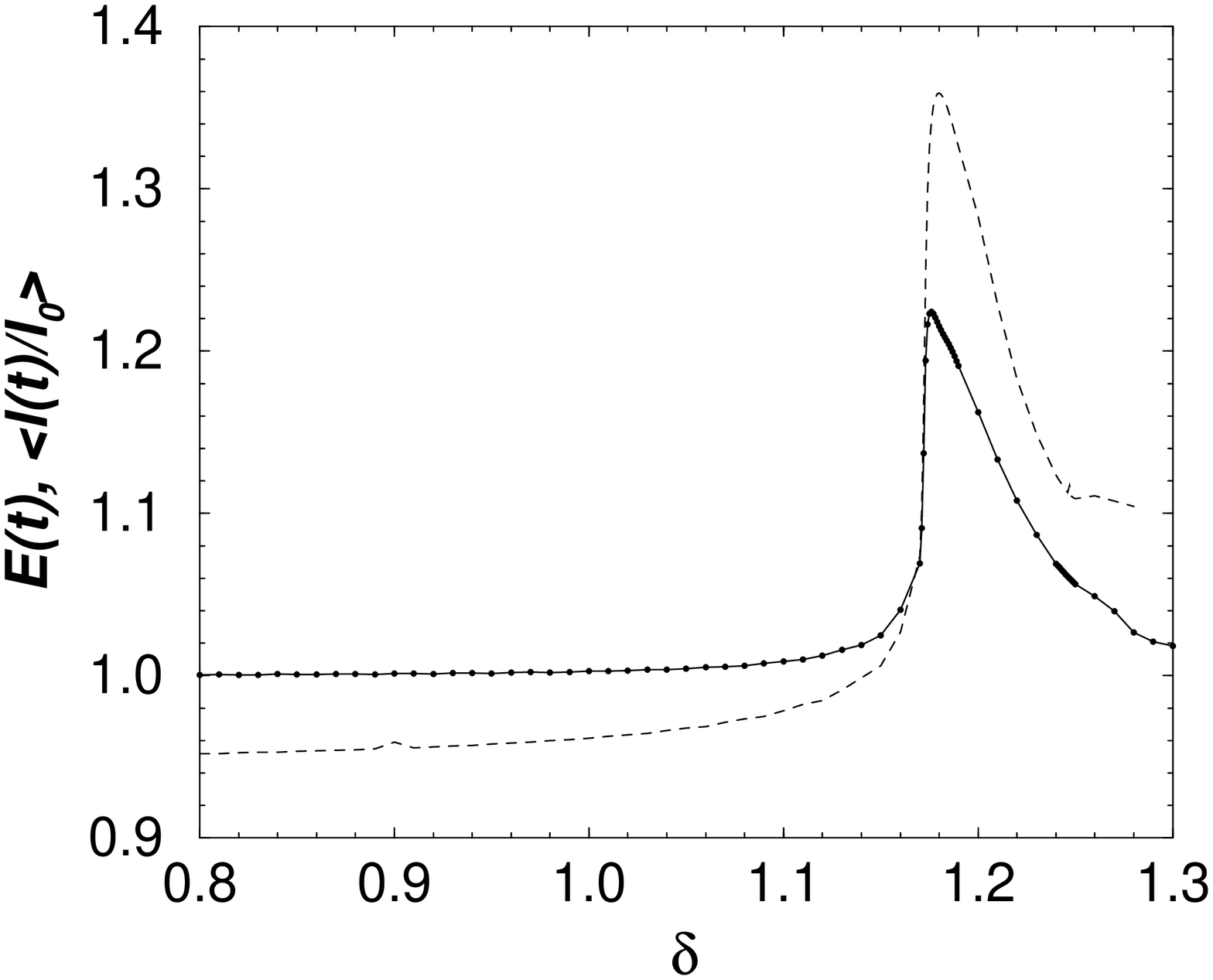, width=2.3in, angle=-90} 
\caption{
$\langle l(t)/l_{0} \rangle$ (points connected by  
solid line) and 
$\langle E(t) \rangle$ (dashed line) vs $\delta$. 
Upper panel: Collective coordinate approach. 
The resonance is at $\Omega_{R}=1.2452$. 
Lower panel: Results from the full system, Eq.\ (\ref{ecua1}). 
The resonance is at $\delta=1.18$, close to $\Omega_{i}=1.2247$.
}
\label{reso}
\end{figure}

In order to present in a more evident manner
the resonance, we have plotted in Fig.\ 
\ref{reso} the values of the mean normalized
width and the mean energy as functions 
of $\delta$. We compute these magnitudes as time averages of 
$l(t)/l_{0}$ and $E(t)$ for times $100 \le t \le 200$, as 
to avoid possible transient effects (which are mostly relevant in the
damped case discussed below). 
We see that on the collective coordinate level
$\langle l(t)/l_{0} \rangle$ and $\langle E(t) \rangle$ 
have a maximum at $\delta = \Omega_{R}$; correspondingly,
in the original system, Eq.\ (\ref{ecua1}), 
the resonance takes place at $\delta=1.18$, close to $\Omega_{i}=1.2247$. 
We have also found that for $1.3 < \delta < \omega_{0}$ (not shown), 
the energy is an increasing function of $\delta$ 
whereas the width of the kink is practically 
constant. This implies that the increment of the
energy is due to excitation of 
the phonons, which becomes more efficient as $\delta\to\omega_0$, 
and is not related with the internal mode. 

\begin{figure}
\epsfig{file=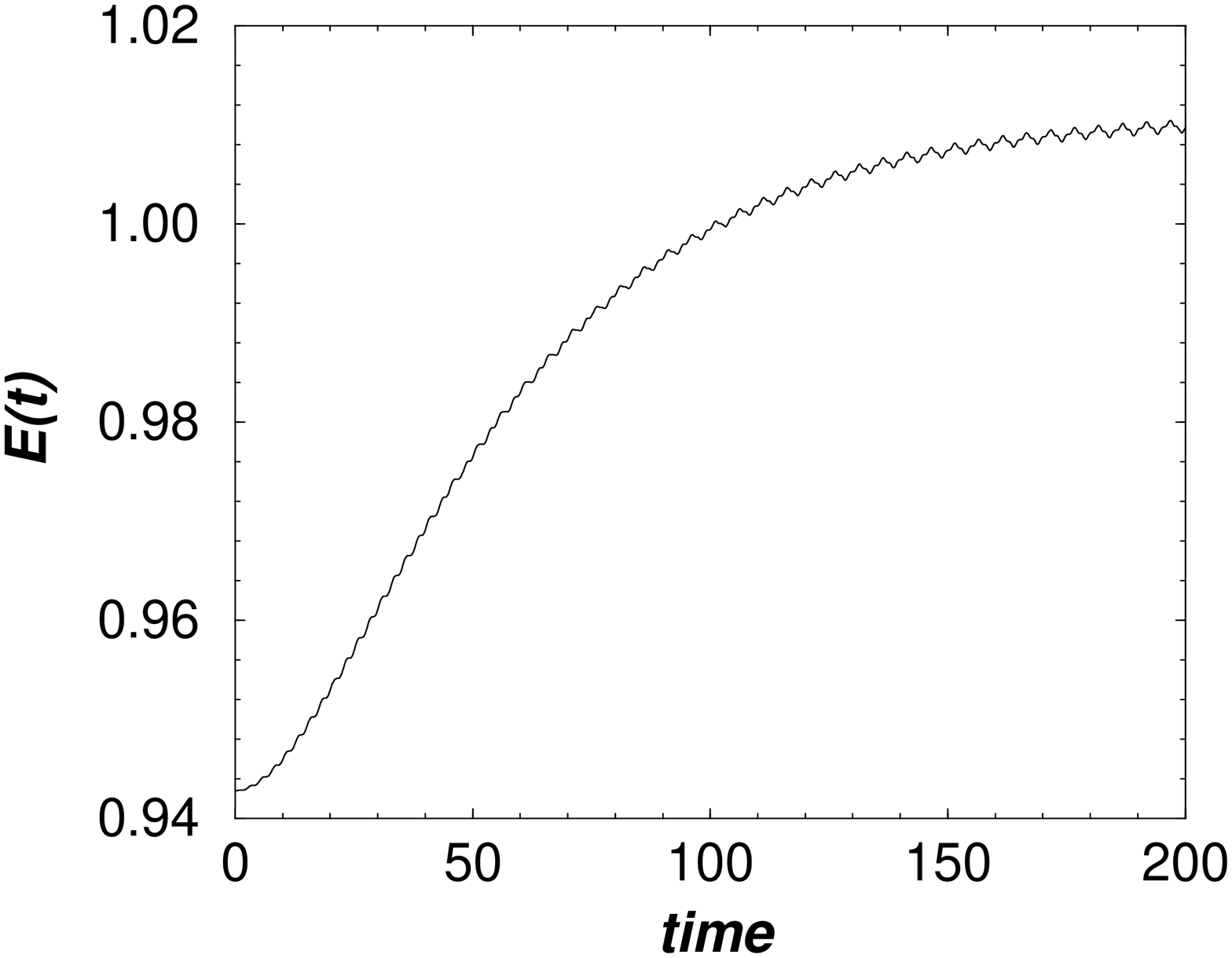,width=2.3in,angle=-90} \\
\\
\epsfig{file=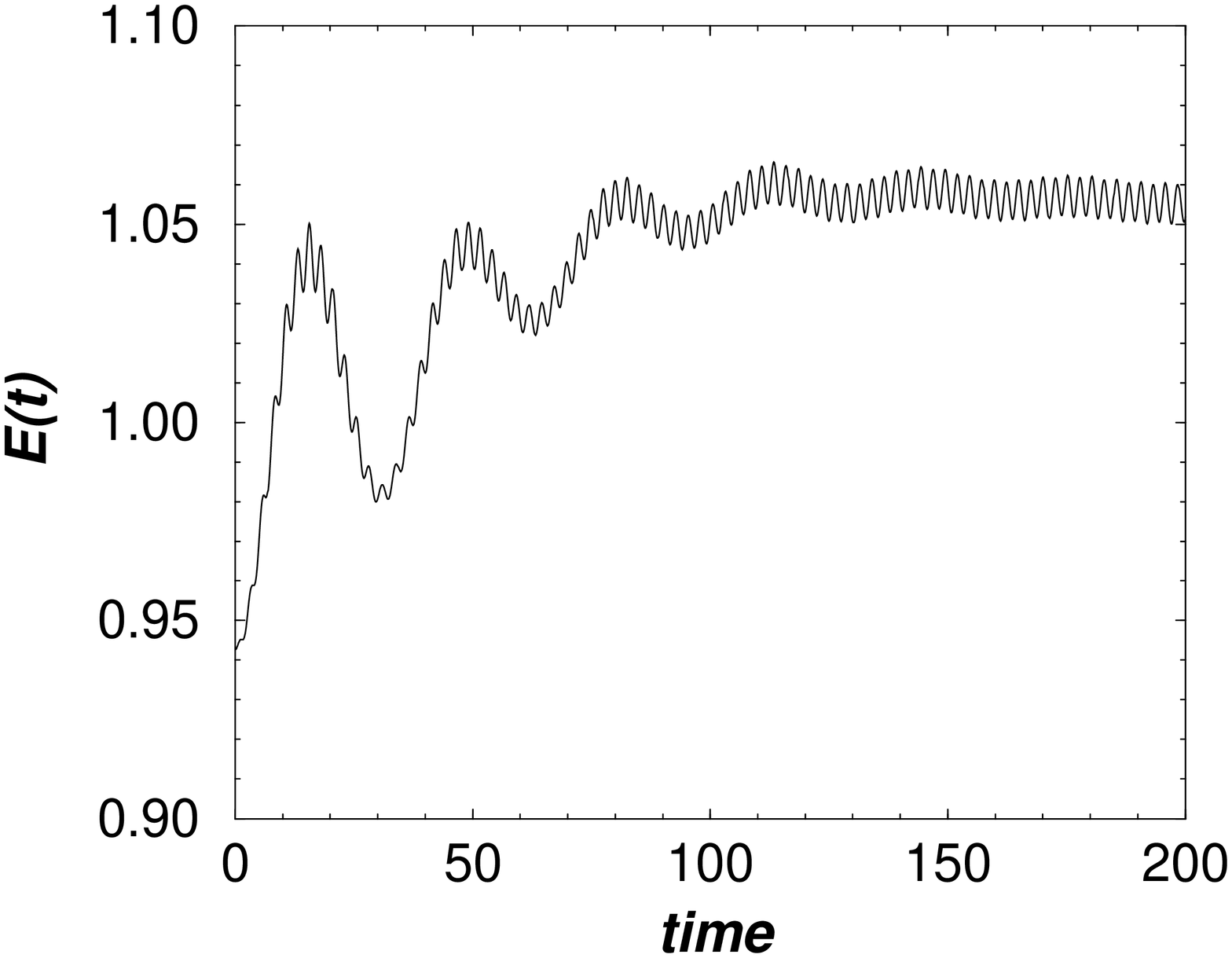,width=2.3in,angle=-90} 
\caption{Damped case, $\beta=0.05$:
Evolution of the energy according to the collective coordinate approach,
obtained by numerical integration of
Eq.\ (\ref{ecua24}), and from the full system, by numerical 
simulation of Eq.\ (\ref{ecua1}),
when $\delta=1.245 \approx \Omega_{R}$ and 
$\delta=1.22 \approx \Omega_{i}$ respectively. 
}
\label{f7y8}
\end{figure}

\begin{figure}
\epsfig{file=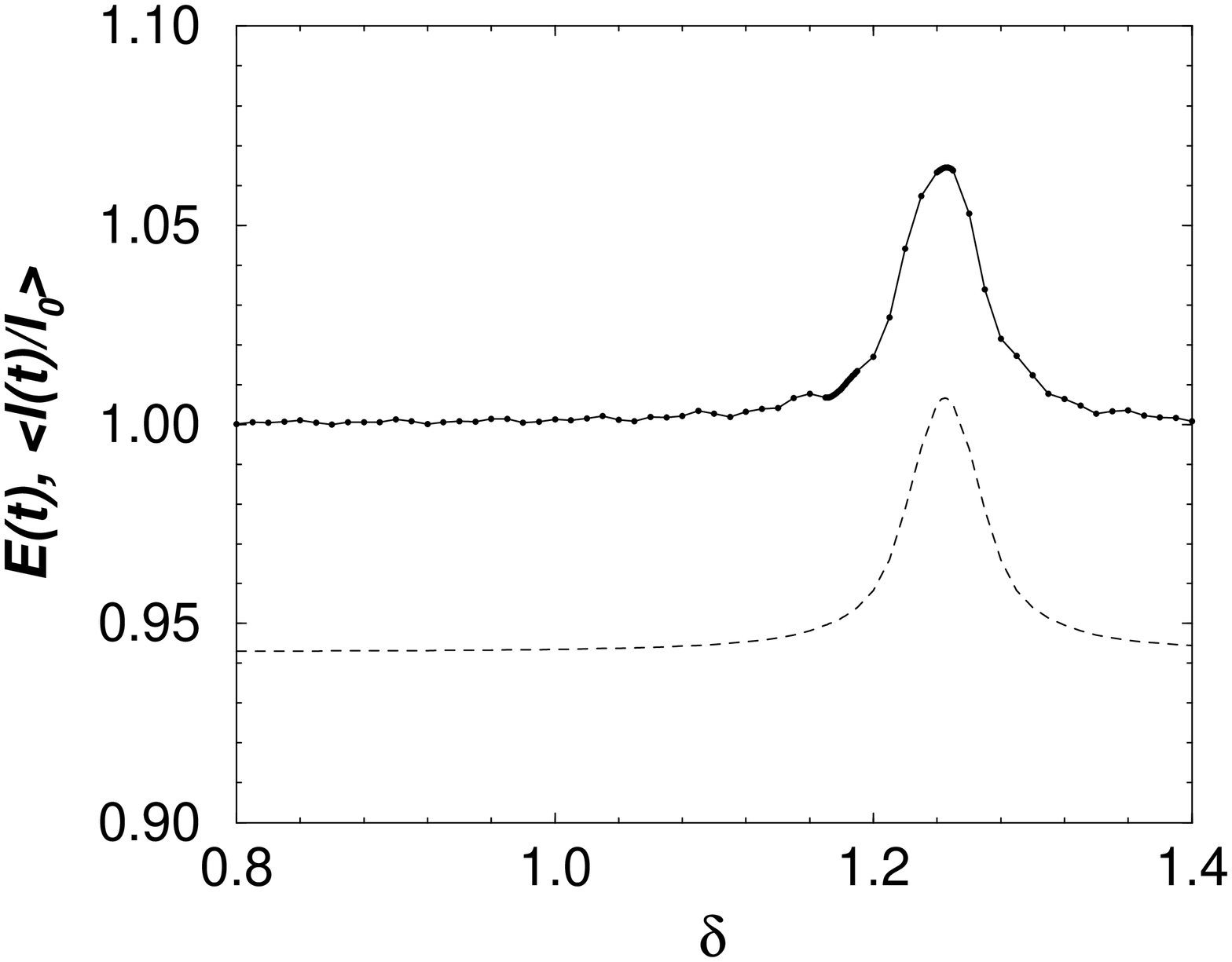,width=2.3in,angle=-90} \\
\epsfig{file=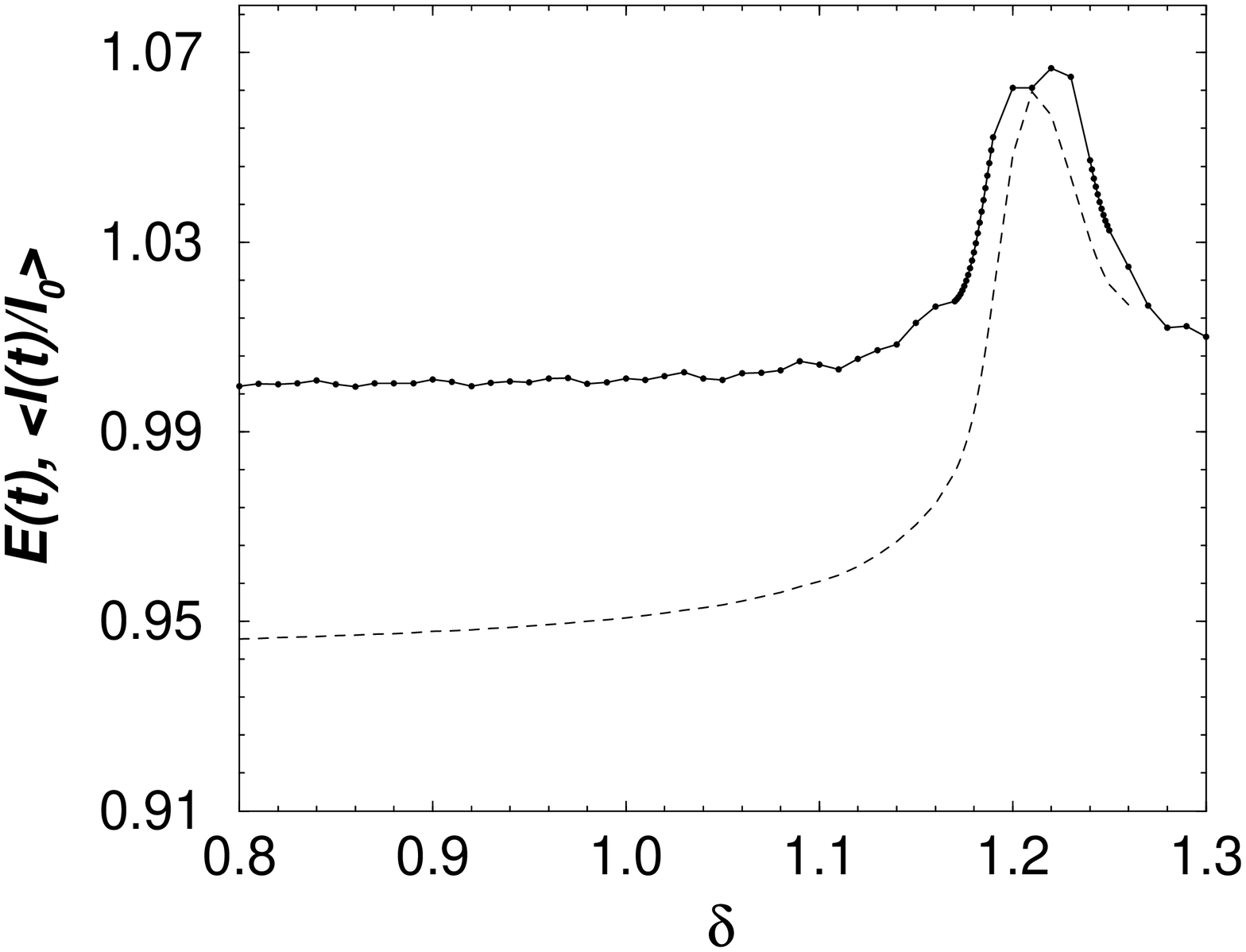,width=2.3in,angle=-90} \\
\caption{Damped case, $\beta=0.05$:
$\langle l(t)/l_{0} \rangle$ (points connected by  
solid line) and 
$\langle E(t) \rangle$ (dashed line) vs $\delta$. 
Upper panel: Collective coordinate approach. 
The width and the energy take their maximum values at $\delta=1.246$ and 
$\delta=1.245$ respectively.   
Lower panel: Results from the full system, Eq.\ (\ref{ecua1}). 
The effective width has its maximum at $\delta=1.22$ and 
the maximum of the energy occurs at $\delta=1.21$, both 
of them close to $\Omega_{i}=1.2247$.
}
\label{resob}
\end{figure}

Coming now to the damped case, as we announced in the preceding section,
we have not obtained any analytical result.
For this reason, in this section we carry out numerical simulations
of Eq.\ (\ref{ecua1}) and also integrate numerically the collective 
coordinate prediction,
Eq.\ (\ref{ecua24}), when $\beta \ne 0$. In Fig.\ 
\ref{f7y8} we plot the evolution of the energy at resonance 
for $\beta=0.05$, as given by the collective coordinates and as 
obtained from the numerical simulations of Eq.\ (\ref{ecua1}).
For each case, we have chosen a frequency near to 
$\Omega_{R}$ (for the collective coordinates) and
$\Omega_{i}$ [for Eq.\ (\ref{ecua1})]. 
To complete the study of the damped case,
we present the result for the resonance as we did for the undamped
case, by plotting the mean width and energy vs $\delta$ for 
$\beta=0.05$ (see Fig.\ \ref{resob}). 
We can observe that,
at resonance, these functions also have maximums, although in the
presence of damping they are less peaked than in the undamped case:
$\langle l(t)/l_{0} \rangle$ increases around a $7\%$ and the energy 
around a $10\%$. Interestingly, the resonance for the full system
is now much closer to the internal mode than in the undamped case. 
We believe that the fact that the damping suppresses quite effectively
the phonons is the reason why the damped system follows more closely
the collective coordinate prediction, which indeed neglects all 
phonon contributions. 
 
\section{Conclusions}
\label{conclu}

In the present paper we have studed the effect of a parametric  
ac force on the dynamics of $\phi^4$ kinks 
with and without damping. By using a 
collective coordinate approach, we have found 
resonances related with the excitation of the internal mode 
when the frequency of the ac force is close to the Rice 
frequency $\Omega_{R} \approx 
\Omega_{i}$. We have verified numerically this prediction by 
computing the solution of the Eq.\ (\ref{ecua1}), its corresponding 
energy and the width of the kink. When this kind of resonance occurs, 
the energy and the width of the system increase, while for other
values of the frequency (except for smaller, secondary resonances) 
they are 
bounded (in some cases periodic) functions. Concerning the mobility of 
the kink, it is very important to point out, at least for the time 
considered here, that all the input energy 
goes to the internal and phonon modes and for this reason the kink 
does not move (neither near nor away from the resonance).    

Although this work has focused on the $\phi^4$ kink, experience 
shows that it is very likely that similar phenomenologies will be
observed in the dynamics of other solitary waves with internal modes. 
In this respect, it is important to 
compare the above results to those obtained for 
non-parametric ac drivings \cite{prl} in order to give a thorough,
coherent picture of the problem. The collective coordinate analysis,
confirmed by the direct numerical simulations of the full system,
shows that when we drive the 
$\phi^4$ system parametrically 
the force acts directly and non-parametrically on the width of the kink,
i.e., the internal mode is excited and the phenomenon
of resonance takes place and can be observed by monitoring 
$l(t)$ and $E(t)$. On the contrary, non-parametric drivings affect 
the internal mode only indirectly through the coupling to the 
translation mode, which in turn leads to a parametric coupling. 
We thus see that the internal mode dynamics is responsible for a
highly non-trivial behavior: non-parametric drivings on the full
system induce parametrical resonances in the kink dynamics, and 
vice versa. As a final remark, we want to point out that our results
indicate a way of exciting internal modes without affecting the 
translation mode, which can mask the effects one is interested in. 
As we have seen, the parametric ac force affects only the internal
mode, and the kink remains at its original position (or moving 
uniformly with the same speed if nonzero) at all times. This is 
not possible with a non-parametric driving and might be a mechanism
of interest in situations where the role of internal modes must
be elucidated.

\section{Acknowledgments}

Work at Sevilla has been supported by the EU grant
LOCNET HPRN-CT-1999-00163.
Work at Legan\'es has been supported by 
DGI of MCyT (Spain) through grant BFM2000-0006.
Travel between Spain and Bayreuth has been
supported by the Emil-Warburg-Foundation of 
Universit\"at Bayreuth.

\end{multicols}

\twocolumn

\end{document}